# 21st Century Global and Regional Surface Temperature Projections

Nicole Ma[1], Jonathan H. Jiang[2], Kennard Hou[3], Yun Lin[4], Trung Vu[4], Philip E. Rosen[5], Yu Gu[4], Kristen A. Fahy[2]

1. Sage Hill School, Newport Beach, CA 92657
2. Jet Propulsion Laboratory, California Institute of Technology, Pasadena, CA 91109
3. Hanford High School, Richland, WA 99354
4. JIFRESSE, University of California, Los Angeles, CA 90095
5. Independent Researcher, Vancouver, WA 98662

Correspondence: Jonathan.H.Jiang@jpl.nasa.gov

**Abstract**

Many countries and regions across the globe broke their surface temperature records in recent years. Recent crippling heat waves swept across the Earth, further sparking concerns about the impending arrival of "tipping points" later in the 21st century. This study analyzes observed global surface temperature trends in three target latitudinal regions: the Arctic Circle, Tropics, and the Antarctic Circle. We show that global warming is accelerating unevenly across the planet, with the Arctic warming at more than three times the average rate of our world. We also analyzed the reliability of latitude-dependent surface temperature simulations from a suite of Coupled Model Intercomparison Project Phase 6 (CMIP6) models and their multi-model mean (MMM) by comparing their outputs to observational datasets. We selected the best-performing models based on their statistical abilities to reproduce historical, latitude-dependent values adapted from these datasets. The surface temperature projections were calculated from ensemble simulations of the Shared Socioeconomic Pathway 2-4.5 (SSP2-4.5) by the selected CMIP6 models. We determine the timing when the climate will warm 1.5, 2.0, and 2.5 °C relative to the preindustrial period globally and in the three target regions. Our results reaffirm a dramatic, upward trend in projected climate warming acceleration, with unprecedented exponential growth in the Arctic, which could lead to catastrophic consequences across the Earth. Further studies are necessary to determine the most efficient solutions to reduce global warming acceleration and maintain a low SSP, both globally and regionally.

## 1 Introduction

In 2022, crippling heat waves swept throughout the globe, taking a dramatic, deadly turn as they killed thousands and gave rise to devasting continent-wide wildfires (Gillespie *et al*., 2022; Mustafa, 2022; Reese, 2022). At least 90 people were killed in India and Pakistan alone due to a record-shattering heat wave in late March that soon led to one of the hottest March-April periods in South Asian history, and temperatures in Pakistan even reached up to a scorching 49.5 ºC in May. A record 47.0 ºC daily high was recorded in Portugal (Kwon, 2022), and heat-induced complications greatly contributed to the 53,000 excess deaths across the European continent in July (Mandiá, 2022).

Historically, fatal heatwaves have been linked to anthropogenic global warming (Mitchell *et al.*, 2016). Hansen *et al.,* (2020) reaffirmed the prevalence of global warming acceleration in the past half decade through the large deviation of global temperature anomalies from the linear warming rate of 1970-2020. The study attributes this pronounced acceleration to the energy imbalance of our planet and an increase in net climate forcing. Accelerated global warming has had substantial impacts on the global hydrologic cycle, food production, energy, health, natural disasters, and socioeconomics (Gou *et al.,* 2020; Asseng *et al.,* 2015; Mcglade and Ekins, 2015; Colón-González *et al.,* 2018; Diffenbaugh *et al.,* 2017; Burke *et al.,* 2015).

However, global warming is not uniform across the planet. In recent decades, the Arctic has been warming at more than three times the global mean (AMAP, 2020), and the region's surface temperatures demonstrate an exponential growth that many of the world's foremost climate models fail to replicate. Continued acceleration will likely amplify the positive ice-albedo feedback loop and lead to disastrous polar vortexes in the Northern Hemisphere (NH, Kretschmer *et al.*, 2018) and worldwide sea level rises and flooding (Box *et al.,* 2018). Latitudes in the equatorial region and Southern Hemisphere (SH, i.e., the Equator, Southern Temperate, and Antarctic Circle) have demonstrated slower warming trends than higher latitudes (Gleisner *et al.,* 2020). Although sea-ice loss is noted in both the Arctic and Antarctic Circles, the Antarctic Circle has a warming acceleration observed clearly lower than the global mean. The Tropics, Southern Temperate, and Antarctic Circle maintain linear trends with a low rate of warming.

As global warming continues to accelerate, warming-induced tipping points become increasingly imminent. A tipping point in the climate system is a critical threshold that leads to large, irreversible changes when crossed. The crossing of a tipping point is likely at 1.5 °C warming above pre-industrial levels and highly probable at 2 °C (Lenton *et al.*, 2019). The 2021 NOAA (National Oceanic and Atmospheric Administration) Global Climate report warns that surface temperatures in 2021 have already reached 1.04 °C of warming above levels in the pre-industrial period (NOAA, 2021). According to the 2018 IPCC special report on the impacts of global warming at 1.5 °C, temperature increases above this threshold will lead to unprecedented changes in all facets of life — from the increased frequency of extreme events to food insecurity to a significant decrease in biodiversity (IPCC, 2018). Warming at this level is projected to lead to glaciers melting in the high mountains of Asia (Yao *et al.*, 2012; Kraaijenbrink *et al.,* 2017). Warming of 2 °C will have even more disastrous results; in the agricultural industry alone, it will lead to substantially smaller yields of essential crops such as maize and rice, especially in sub-Saharan Africa, Southeast Asia, and the Americas, compounding the issue of food security (Mbow *et al.*, 2019). Crossing this tipping point is projected to lead to an average global ocean rise of 20 cm (Jevrejeva *et al.*, 2016), and large increases in extreme storms, drought, and fire weather in numerous regions (IPCC, 2021). Different regions will reach these temperature thresholds at different times due to the aforementioned variations in warming rates. Because of differences in warming rates across different areas and the approaching perils as surface temperatures continue to accelerate, it is becoming increasingly important to understand what the future of surface temperatures will look like — not just globally, but also in different latitudinal regions.

GCMs (Global Climate Models), which simulate climate systems on both regional and global scales, are imperative in understanding climate variability and change (IPCC, 2013), and therefore are a crucial tool in predicting the future of surface temperatures. However, it is well known that GCMs have large variabilities in simulating historical surface temperatures and some models have failed to reproduce historical temperature trends (Knutti *et al.*, 2017; Liang *et al.*, 2020). Recently, a more advanced wave of GCMs were used in the Coupled Model Intercomparison Project Phase 6 (CMIP6) with more extensive physical simulations and higher spatial resolution (Eyring *et al.*, 2016).

Papalexiou *et al.* (2020) analyzed the accuracy and reliability of CMIP6 global mean temperature simulations based on various criteria such as trend conformity and autocorrelation. The team found that all models showed differences in distributional shapes when compared to historical data and there were significantly varied performances between models across all metrics. The study also revealed that although models are able to reproduce historical trends, many simulations fall short in other areas such as long-term potentiation (LTP) and distributional shapes. Additionally, Fan *et al.* (2020) analyzed mean and extreme surface air temperatures produced by 16 different CMIP6 models. The study found that although the majority of models are able to accurately reproduce spatial patterns of global climatological mean temperatures, there still exists a large spread across different models and regions. Drastic differences in climate sensitivities and warming rates in different regions of the planet underscore the need for closer analyses of climate models' latitude-dependent robustness, which will in turn reveal patterns in regional surface temperature acceleration.

In this study, we first analyze the global and regional warming trends based on observations, including GISS (Goddard Institute for Space Studies) Surface Temperature (GISTEMP) and Berkeley Earth temperature datasets. We substantiate trend variance amongst latitudes and analyze their historical behaviors. Then, we examine the reliability of CMIP6 models and project $21^{st}$ century global and regional surface temperatures to foster a better understanding of future surface temperatures and be better equipped to predict when tipping points will be reached. The selected CMIP6 models with best performance and their multi-model mean (MMM) were evaluated by comparing latitude-dependent surface temperature warming rates with observational data from the two sources. The MMM was considered because no single model excels in every metric (Papalexiou *et al.*, 2020; Fan *et al.*, 2020), especially when analyzed regionally. The best models were then selected and ensemble members were averaged with an associated margin of error to project $21^{st}$ century surface temperatures, both globally and in three regions – the Arctic, Antarctica, and the Tropics – to estimate the arrival of tipping points. The results are important for effectively organizing and prioritizing different regions in the battle against global warming.

## 2 Data and Methodology

### 2.1 Observational Surface Temperature Data

Two observational datasets were considered:

(a) The GISS Surface Temperature Analysis (https://data.giss.nasa.gov/gistemp) Land-Ocean Temperature Index (L-OTI) on a regular 2º × 2º grid. The L-OTI is compiled by combining NOAA GHCN v4 SAT (Global Historical Climatology Network version 4 Surface Air Temperature) anomalies over land and sea ice with ERSST v5 (Extended Reconstructed Sea Surface Temperature version 5) oceanic data over water (Hansen *et al.*, 2010). GISTEMP is widely used to observe regional temperature trends and has greater polar coverage than most other datasets.

(b) The Monthly Land + Ocean Average Temperature Dataset from Berkeley Earth (http://berkeleyearth.org/data/), calculated by averaging air temperatures at sea ice. The combined dataset was created by concatenating Berkeley Earth land data with a spatially kriged version of HadSST3 (Hadley Centre Sea-Surface Temperature dataset version 3) (Rohde *et al.*, 2020).

We converted monthly data to annual data using resampling techniques. Berkeley Earth data was linearly interpolated from a 1º × 1º grid onto a regular 2º × 2º grid, and only data from 1975-2021 was considered. To observe trends in our target regions, we spliced data latitudinally using the following boundaries: the Arctic Circle (66ºN-90ºN), the Antarctic Circle (90ºS-66ºS), and the tropics (24ºS-24ºN). To compare warming rates and verify the prevalence of acceleration in different regions, ordinary least-squares was used to compare the 60-month moving averages of regional surface temperatures to global surface temperatures.

### 2.2 CMIP6 Data

Surface temperature projections from CMIP6 models were accessed through the Earth System Grid Federation (ESGF). All available variants were used for each model and simulation outputs were linearly interpolated onto a regular 2º × 2º grid. To compare CMIP6 models in terms of their historical simulation performance, we resampled the time coordinate to an annual frequency. We sliced historical data to the years 1975-2014, and SSP2-4.5 projections were averaged annually for the years 2015-2099 for all selected models. When compared to historical data, the base period of 1951-1980 was used for conformity with observed datasets. Margins of error were calculated based on nuances in the realizations of both historical and forecasted data. For estimating the modeled future temperature projection, we subtracted the average of surface temperatures using the pre-industrial period 1880-1920 for each historical simulation of the model. Differences in the anomalies were the main consideration for quantifying margins of error. Details regarding each of the 11 models are illustrated in Table 1.

**Table 1** – List of CMIP6 Models and Properties.

| # | Model | Region | Institution |
|---|---|---|---|
| 1 | AWI-CM-1-1-MR | Germany | Alfred Wegener Institute, Helmholtz Centre for Polar and Marine Research |
| 2 | BCC-CSM2-MR | China | Beijing Climate Center |
| 3 | CESM2 | USA | National Science Foundation |
| 4 | E3SM-1-1 | USA | Department of Energy |
| 5 | EC-Earth3 | Europe | Europe-wide Consortium |
| 6 | FGOALS-g3 | China | Chinese Academy of Sciences |
| 7 | GFDL-CM4 | USA | National Oceanic and Atmospheric Administration, Geophysical Fluid Dynamics Laboratory |
| 8 | GISS-E2-1-G | USA | Goddard Institute for Space Studies |
| 9 | IPSL-CM6A-LR | France | Institut Pierre Simon Laplace |
| 10 | KIOST-ESM | Korea | Korea Institute of Ocean Science and Technology |
| 11 | NorESM2-LM | Norway | NorESM Climate Modeling Consortium |

### 2.3 Methods

In order to create quantitative latitude-dependent analysis, we examined the linear trends of models and observational datasets from 1975-2014 at each latitude band, averaged out over all respective longitudes. Ordinary, least-square is the standard approach for linear regression analysis, which minimizes the sum of the squares of the differences between the dependent variable and the linear function of the independent variable. For each latitude, we calculate the slope *b* of surface temperature linear trends through least-squares regression using the following equation:

$$b = \frac{\Sigma(x_i - \overline{x})(y_i - \overline{y})}{\Sigma(x_i - \overline{x})^2}$$

In this equation, $x_i$ = value of the independent variable (latitude) at the *i*th data point, $y_i$ = value of the dependent variable (surface temperature) at the *i*th data point, $\overline{x}$ = mean value of the independent variable, and $\overline{y}$ = mean value of the dependent variable.

Three metrics were then used to compare latitude-dependent model simulations with observational data to determine the most reliable CMIP6 models: (1) the coefficient of determination ($r^2$, a statistically-derived measure of agreement between actual data and corresponding modeled results), where values of $r^2$ closer to 1 indicate a stronger performance; (2) the root mean square error (RMSE), where the best-performing model simulations with respect to RMSE have values close to 0; (3) the mean squared error (MSE), where the best performing model simulations also have values close to 0.

Furthermore, we constructed Taylor diagrams to graphically analyze the models' conformities with observational data (Taylor, 2001). Taylor diagrams are advantageous when determining the relative performance of several different models (IPCC, 2001) and they quantify degrees of correspondence between modeled and observed behavior using RMSE, as well as two additional metrics: (4) the Pearson correlation coefficient (*r'*) and (5) the standard deviation (*s*). The Pearson correlation measures the linear relationship between two sets of data — observational and modeled. The sign of the obtained value indicates the positive/negative correlation, and the magnitude of *r'* indicates the strength of that correlation. The standard deviation is a widely used measure of the variation of a set of values and is calculated by the square root of the sample's variance. The standard deviation was also used to quantify margins of error for tipping point estimations.

## 3 Results
### 3.1 Observational Analysis

Figure 1 depicts the 60-month moving average of surface temperature anomalies from 1975 to 2021 with the base period of 1951-1980, as well as their linear trends. Polar regions have significantly more intra-annual variation than the planet as a whole, while tropical surface temperatures have less variation both intra-annually and interannually. Berkeley Earth data estimates more dramatic temperature extremes for the polar regions, but the two datasets generally corroborate each other in terms of linear trends. GISTEMP estimates a higher warming rate in all regions than does Berkeley Earth data except for the Arctic, where the trend obtained through the GISTEMP datasets differs by approximately 2.32% from the trend obtained through Berkeley Earth data. The greatest disagreement occurs in the tropics, where the GISTEMP trend differs from the Berkeley Earth trend by 31.9%. These differences are accounted for during the evaluation of CMIP6 models in comparison to historical data.

Based on the 60-month moving average of global and arctic surface temperatures from 1975-2021, the Arctic is warming at around 3 times the rate of the planet. This result is consistent with the estimate from the Arctic Monitoring and Assessment Programme (2021). Significant deviations from the linear trend are attributed to unforced ENSO (El Niño-Southern Oscillation) variability. The difference in warming rates between the Arctic region and the planet is explained through Arctic polar amplification (Lee, 2014), which can be attributed to various factors such as the ice-albedo feedback loop (Petoukhov and Semenov, 2010; Screen, 2013; Tang *et al.*, 2013), changing oceanic currents (Lee, 2012; 2014), and polar jet streams (Francis and Vavrus, 2012).

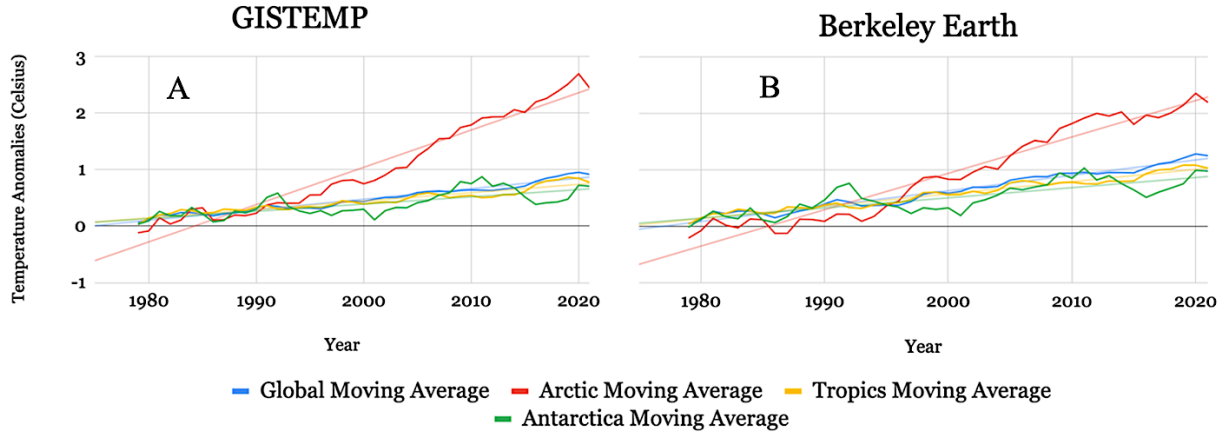

**Figure 1**: (A) GISTEMP and (B) Berkeley Earth Regional Surface Temperature Anomalies from the years 1975-2021, with respect to the 1951-1980 base period.

The Antarctica is warming at slightly more than half the rate of the planet (0.56 times), largely due to high ice sheet orography (Singh *et al*., 2020), weaker Antarctic surface albedo feedback, ocean heat uptake in the Southern Ocean, Antarctic ozone depletion (Masson-Delmotte *et al*., 2013), and varying Antarctic surface heights (Salzmann, 2017). There is also significant intra-annual variation, begetting a low coefficient of determination with respect to the Antarctic 60-month moving average.

Tropical surface temperatures yield a linear trend with the highest conformity to the warming rate of the planet, illustrated by its near linear warming rate of 0.86 times the planet. The Tropics have lower intra-annual temperature variation when compared with latitudinal extremes and follow the trend of the overall planet more closely. This is explained by radiative resistance to temperature change caused by tropical tropospheric temperature variations (Spencer *et al*., 2019).

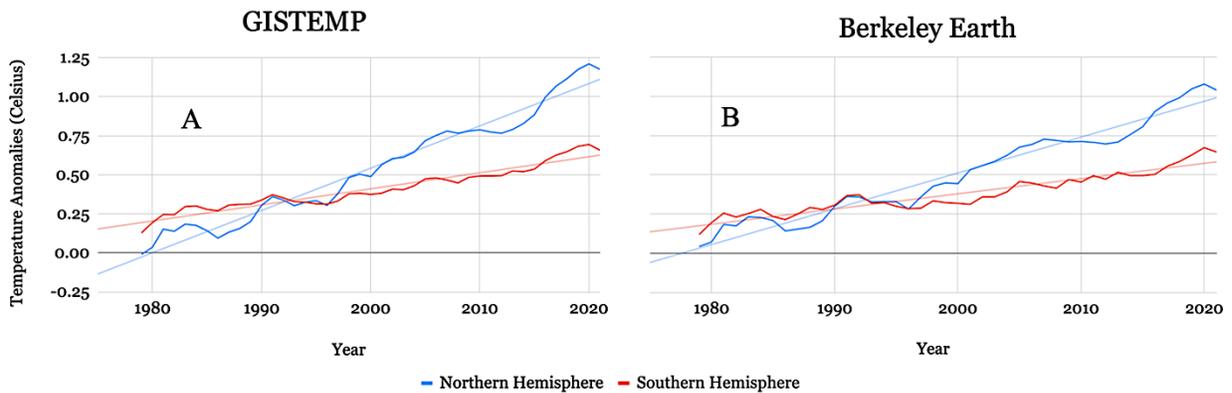

**Figure 2**: Northern Vs. Southern Hemispheric Surface Temperature Anomalies.

Figure 2 reaffirms the significant difference in warming rates between the NH and the SH, with the Northern Hemisphere warming at 2.62 times the rate of the Southern Hemisphere. The discrepancy in warming rates between the two hemispheres is largely due to the higher amount of landmass in the Northern Hemisphere, the cross-equatorial Atlantic Ocean heat transport, and the albedo difference between Antarctica and the Arctic (Feulner *et al.*, 2013; Kang *et al.*, 2015). As the north warms at a faster rate, equatorial tropical rain bands shift northward, consequentially drying out the Southern Hemisphere. If this trend continues and the temperature difference between the two hemispheres continue to enlarge, it could lead to significant alteration of tropical rainfall patterns, affecting all corners of the world (Friedman *et al.*, 2013).

### 3.2 CMIP6 Model Evaluation

Figure 3 illustrates the latitude-dependent warming rates of each of the 11 CMIP6 models, as well as the MMM. While most models are able to reproduce general spatial patterns of surface temperatures, there remain substantial deviations and variation amongst the models. Furthermore, although all models and the MMM conform to the trends in the tropical region, there is a significant spread in mid-latitudinal and polar regions. The MMM performs, on average, better than most of the individual models.

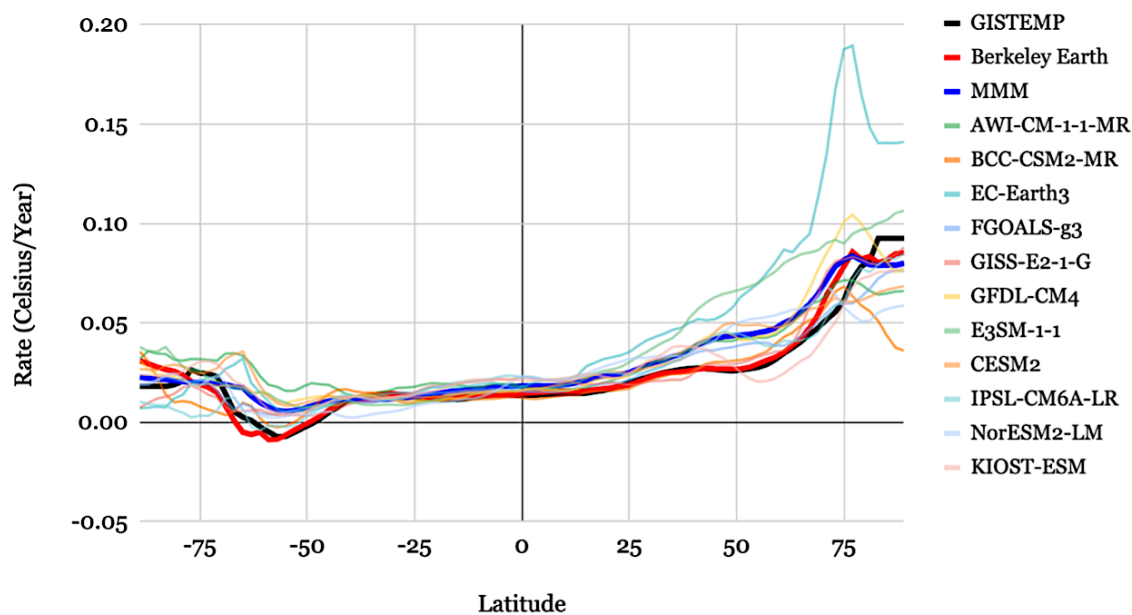

**Figure 3**: Latitude-dependent analysis of CMIP6 model; the rate of warming across latitudes from 1975-2014.

The global scale Taylor diagrams shown in Figure 4 support the conclusions from the tables (see Table S1 and S2 in the Supplementary: Supporting Information section). EC-Earth3 deviates significantly from the reference dataset and the cluster of models, and E3SM-1-1 also differs from the other models. GISS-E2-1-G performs the best in both diagrams in terms of its conformity to the standard deviations of the reference datasets. The RMSE of each model is signified by the distance from the point representing the model to the reference point on the x-axis. Despite having more deviance in terms of its standard deviation, FGOALS-g3 has a lower relative RMSE than GISS-E2-1-G in Figure 4A. Regional scale Taylor diagrams for the Arctic, Tropics, and Antarctic are shown in the supplementary. The diagrams reveal a greater spread in performance between models' regional projections than between their performance in global projections. Most models' Arctic projections have a smaller standard deviation than the observational datasets, suggesting that models generally have difficulty projecting Arctic temperature variability.

Table S1-I in the supplementary section shows the statistical indices (RMSE, MSE, and $r^2$) of each latitude-dependent simulation with respect to Berkeley Earth historical data, Table S1-II depicts the indices with respect to GISTEMP data. While there is slight variation in the rankings of the models, the top two positions in both tables are composed of GISS-E2-1-G and FGOALS-g3. E3SM-1-1 and EC-Earth3 consistently perform significantly worse than the other models. The MMM performs better with respect to Berkeley Earth data than GISTEMP data, although it remains in the top five for both comparisons.

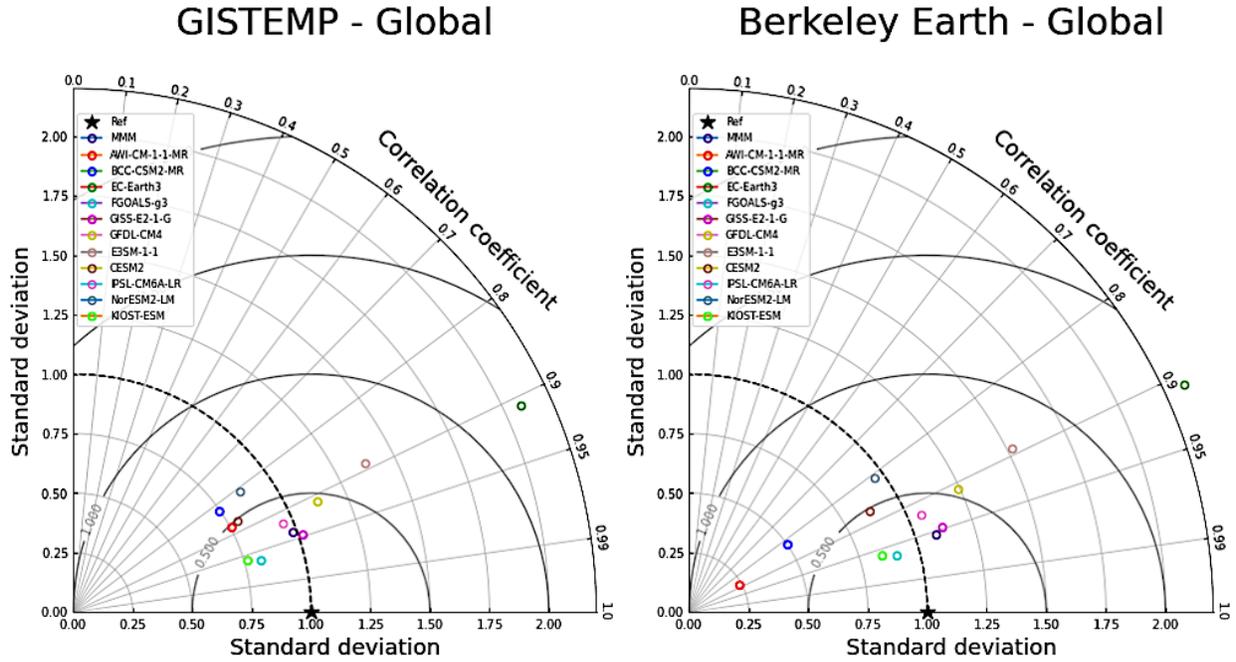

**Figure 4**: Taylor diagrams evaluating CMIP6 models relative to global observational datasets.

### 3.3 Temperature Projections

Tables S2-I through S2-VIII in the supplementary section list estimations of the calendar years in which the 1.5 °C, 2.0 °C, and 2.5 °C tipping points will be reached globally and in the three target regions. Only models with more than one realization for SSP2-4.5 were used to estimate tipping points, although projections from all 11 models were used to calculate the MMM. In all models and the MMM, the 1.5 °C, 2.0 °C, and 2.5 °C thresholds in the Arctic have already been passed prior to 2014. Models such as CESM2 have significant deviations from FGOALS-g3 and GISS-E2-1-G in the Antarctic, which can be attributed to a lack of simulated dynamic ice sheets in the region. FGOALS-g3 and GISS-E2-1-G, the two models deemed most reliable by the indices in the previous section, generally agree with each other in terms of the estimation of the arrival of tipping points. However, there exists a large spread amongst the models altogether, especially in the highest temperature threshold of 2.5 °C.

Table S3 in the supplementary section shows the RMSEs for all CMIP6 models against regional observations. As corroborated by Figures S1-3, there exists a greater spread between model performances regionally when compared to global projections. Some models, such as BCC-CSM2-MR, perform poorly in one region while excelling in another. Although BCC-CSM2-MR performs poorly in terms of Arctic projections, it has the lowest RMSE among all models for tropical projections, with respect to both observational datasets. FGOALS-g3 performs exceptionally well in terms of Antarctic projections when compared to both observational datasets, maintaining a significant margin from other models. Furthermore, similar to its performance when compared with global observational datasets, the MMM has a lower RMSE than most models across all three regions.

Figure 5 shows the regional distributions of simulated tipping points. The tipping points were determined using projections from all 11 models, as well as the MMM. Ranges increase as the degrees of warming increase, along with the difference between the 2$^{nd}$ and 3$^{rd}$ quartiles. To make the calculations more tractable, calendar years greater than 2100 - a period that is beyond the scope for CMIP6 projections - were counted as the year 2100. There is no separate figure for the Arctic region because models unanimously agreed that the tipping point had been passed before 2014. Significant ranges reveal that projection models still do not agree with each other despite being able to reproduce past general trends in surface temperatures.

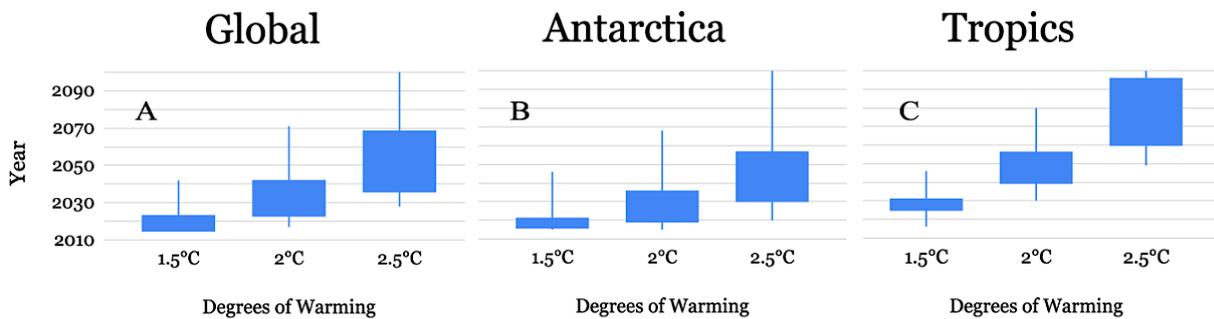

**Figure 5**: Spread of tipping point timing estimations for 1.5, 2.0 and 2.5 °C of warming.

Figure 6 shows projections for the two models deemed most reliable through latitude-dependent analysis: FGOALS-g3 and GISS-E2-1-G. There is a significant difference between the two models' Antarctic surface temperature projections, but the models agree closely in their tropical projections. Based on the projections, the Arctic will continue to be the most severely impacted by global warming, with temperatures estimated to exceed 6 °C above pre-industrial times by the end of the 21$^{st}$ century. The upward concavity of FGOALS-g3 projections in the Arctic suggests continued warming acceleration in the region. This dramatic degree of warming would be expected to have significant impacts globally, especially in regard to polar vortexes and rises in global sea levels. All regions are expected to surpass the 2.0 °C tipping point well before the end of the 21$^{st}$ century, should no countermeasures be taken. All regions, except the tropical projections from FGOALS-g3, are estimated to pass the 2.5 °C threshold by the end of the 21$^{st}$ century.

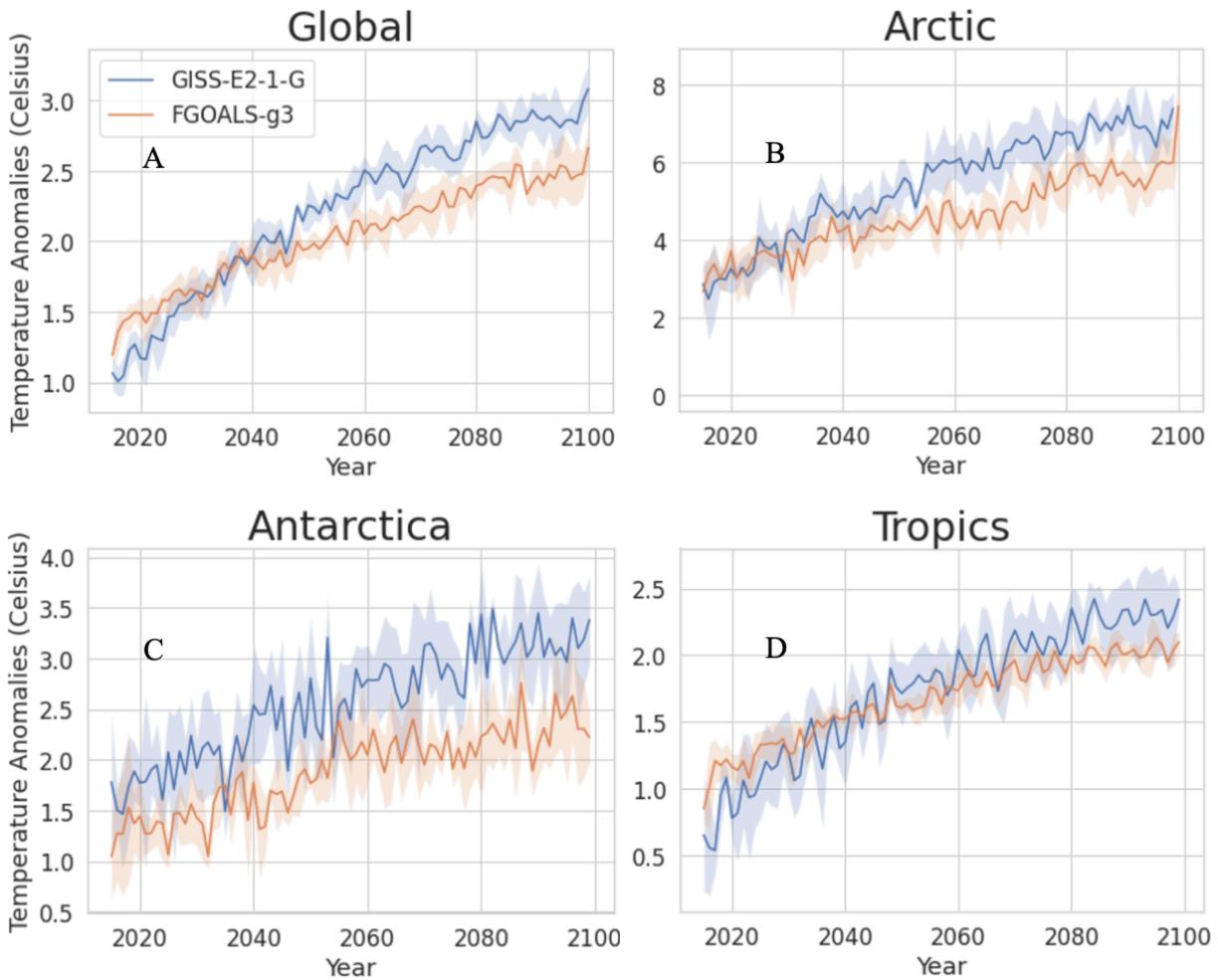

**Figure 6:** 21st century projections for global and regional temperature changes. The shadings denote symmetrical margins of errors.

**4 Conclusions and Discussions**

Two observational datasets and 11 CMIP6 models were used in this study to determine reliable regional surface temperature projections for the 21st century. We analyzed regional observational datasets to determine warming rates relative to the global linear trend, finding that the Antarctic circle and the tropics are warming slower than the planet as a whole while the Arctic warms significantly faster. We confirmed the inter-hemispheric difference in warming rates and found a dramatic positive trend in projected global warming in three target regions, especially in the Arctic.

FGOALS-g3 and GISS-E2-1-G were deemed the most reliable CMIP6 models for surface temperature projections based on statistical indices when evaluating the models' ability to reproduce historical, latitude-dependent values adapted from the observational datasets. Particular acceleration in Arctic temperatures is observed, with the region estimated to reach 6 °C above pre-industrial levels by the end of the 21st century.

Based on analyses of CMIP6 models, there exists a large variability between different realizations of each model, as well as significant spreads between different models. The MMM generally performs better than most individual models, and we identified areas of improvement needed to create more accurate projections in the next phase. Inaccurate projections, especially in polar regions, can often be explained by an insufficiency of dynamic simulations for those areas.

Understanding when tipping points will be reached in different regions, as well as deepening apprehension of global warming acceleration, is a key element in planning solutions to curb this pressing issue. Further research on solutions to mitigate global warming acceleration, as well as analysis on different SSPs both globally and regionally, will be imperative in fighting climate change and ensuring the survival of our planet.

**Acknowledgements:** This study is supported by the Jet Propulsion Laboratory, California Institute of Technology, under contract with NASA. We also acknowledge the supports by the Joint Institute for Regional Earth System Science & Engineering, University of California, Los Angeles.

**Data availability:** All data used for this study can be downloaded from the public websites. The GISS Surface Temperature data are available at https://data.giss.nasa.gov/gistemp/, the Berkeley Earth temperature data are available at https://berkeleyearth.org/data/, and the CMIP6 climate models data are available at https://pcmdi.llnl.gov/CMIP6/. For additional questions regarding the data sharing, please contact the corresponding author at Jonathan.H.Jiang@jpl.nasa.gov.

**Supplementary**

The following tables provides the supporting information discussed in the paper.

*Table S1*: *Statistical indices of CMIP6 models in comparison to observational datasets.*

| (I) Model Performance Respective to Berkeley Earth Data | | | | | (II) Model Performance Respective to GISTEMP Data | | | | |
|---|---|---|---|---|---|---|---|---|---|
| Rank | Model | RMSE | MSE | $r^2$ | Rank | Model | RMSE | MSE | $r^2$ |
| 1 | GISS-E2-1-G | 0.0000454 | 0.00674 | 0.913 | 1 | FGOALS-g3 | 0.0000496 | 0.00704 | 0.899 |
| 2 | FGOALS-g3 | 0.0000571 | 0.00755 | 0.89 | 2 | GISS-E2-1-G | 0.0000591 | 0.00769 | 0.88 |
| 3 | MMM | 0.0000772 | 0.0088 | 0.851 | 3 | KIOST-ESM | 0.0000699 | 0.00836 | 0.858 |
| 4 | IPSL-CM6A-LR | 0.0000918 | 0.00958 | 0.823 | 4 | IPSL-CM6A-LR | 0.0000819 | 0.00905 | 0.834 |
| 5 | KIOST-ESM | 0.0000953 | 0.00976 | 0.817 | 5 | MMM | 0.0000966 | 0.0098 | 0.804 |
| 6 | GFDL-CM4 | 0.00012 | 0.0109 | 0.773 | 6 | CESM2 | 0.000158 | 0.0126 | 0.679 |
| 7 | BCC-CSM2-MR | 0.000127 | 0.0113 | 0.756 | 7 | BCC-CSM2-MR | 0.000165 | 0.0129 | 0.664 |
| 8 | CESM2 | 0.000162 | 0.0127 | 0.69 | 8 | GFDL-CM4 | 0.000166 | 0.0129 | 0.662 |
| 9 | NorESM2-LM | 0.000168 | 0.013 | 0.676 | 9 | NorESM2-LM | 0.000181 | 0.0135 | 0.631 |
| 10 | AWI-CM-1-1-MR | 0.00017 | 0.013 | 0.673 | 10 | AWI-CM-1-1-MR | 0.000184 | 0.0136 | 0.627 |
| 11 | E3SM-1-1 | 0.000367 | 0.0192 | 0.294 | 11 | E3SM-1-1 | 0.000407 | 0.0202 | 0.173 |
| 12 | EC-Earth3 | 0.00109 | 0.033 | –1.1 | 12 | EC-Earth3 | 0.00118 | 0.0344 | –1.403 |

*Table S2*: *Estimation of the arrival timing of tipping points.*

| (I) FGOALS-g3 | | | | (II) GISS-E2-1-G | | | |
|---|---|---|---|---|---|---|---|
| Region | 1.5 °C | 2.0 °C | 2.5 °C | Region | 1.5 °C | 2.0 °C | 2.5 °C |
| Global (Years) | 2024 ± 2.50 | 2054 ± 7.90 | 2087 ± 10.55 | Global (Years) | 2024 ± 1.34 | 2039 ± 2.83 | 2057 ± 5.03 |
| Arctic (Years) | < 2014 | < 2014 | < 2014 | Arctic (Years) | < 2014 | < 2014 | < 2014 |
| Antarctica (Years) | 2020 ± 3.70 | 2031 ± 14.72 | 2060 ± 5.74 | Antarctica (Years) | 2024 ± 1.34 | 2039 ± 2.83 | 2057 ± 5.03 |
| Tropics (Years) | 2030 ± 4.08 | 2069 ± 9.8 | > 2100 | Tropics (Years) | 2031 ± 3.24 | 2052 ± 3.90 | 2086 ± 4.45 |

| (III) IPSL-CM6A-LR | | | | (IV) GFDL-ESM4 | | | |
|---|---|---|---|---|---|---|---|
| Region | 1.5 °C | 2.0 °C | 2.5 °C | Region | 1.5 °C | 2.0 °C | 2.5 °C |
| Global (Years) | 2015 ± 0.97 | 2021 ± 2.64 | 2033 ± 3.26 | Global (Years) | 2032 ± 11.79 | 2051 ± 5.13 | 2076 ± 5.29 |
| Arctic (Years) | < 2014 | < 2014 | < 2014 | Arctic (Years) | < 2014 | < 2014 | < 2014 |
| Antarctica (Years) | 2018 ± 2.92 | 2022 ± 5.53 | 2032 ± 8.72 | Antarctica (Years) | 2016 ± 0.58 | 2025 ± 11.55 | 2050 ± 6.03 |
| Tropics (Years) | 2023 ± 3.75 | 2038 ± 4.85 | 2056 ± 5.07 | Tropics (Years) | 2034 ± 10.39 | 2064 ± 3.51 | 2099 ± 2.12 |

| (V) CESM2 | | | | (VI) NorESM2-LM | | | |
|---|---|---|---|---|---|---|---|
| Region | 1.5 °C | 2.0 °C | 2.5 °C | Region | 1.5 °C | 2.0 °C | 2.5 °C |
| Global (Years) | 2018 ± 4.62 | 2030 ± 2.08 | 2042 ± 3.21 | Global (Years) | 2039 ± 2.31 | 2067 ± 3.51 | 2092 ± 7.55 |
| Arctic (Years) | < 2014 | < 2014 | < 2014 | Arctic (Years) | < 2014 | < 2014 | < 2014 |
| Antarctica (Years) | 2016 ± 0.58 | 2017 ± 2.65 | 2026 ± 4.04 | Antarctica (Years) | 2027 ± 7.02 | 2062 ± 5.03 | > 2100 |
| Tropics (Years) | 2029 ± 3.21 | 2041 ± 2.52 | 2062 ± 5.69 | Tropics (Years) | 2042 ± 2.65 | 2068 ± 1.15 | > 2100 |

| (VII) EC-Earth3 | | | | (VIII) Multi-Model Mean | | | |
|---|---|---|---|---|---|---|---|
| Region | 1.5 °C | 2.0 °C | 2.5 °C | Region | 1.5 °C | 2.0 °C | 2.5 °C |
| Global (Years) | 2016 ± 0.97 | 2025 ± 4.72 | 2041 ± 5.53 | Global (Years) | 2020 ± 8.12 | 2034 ± 14.81 | 2053 ± 21.73 |
| Arctic (Years) | < 2014 | < 2014 | < 2014 | Arctic (Years) | < 2014 | < 2014 | < 2014 |
| Antarctica (Years) | 2018 ± 3.18 | 2030 ± 5.68 | 2044 ± 8.43 | Antarctica (Years) | 2020 ± 6.25 | 2031 ± 16.03 | 2048 ± 25.45 |
| Tropics (Years) | 2026 ± 5.10 | 2045 ± 5.54 | 2066 ± 3.57 | Tropics (Years) | 2028 ± 6.96 | 2050 ± 12.57 | 2075 ± 18.08 |

*Table S3*: *RMSE of regional CMIP6 projections in comparison to observational datasets.*

| (I) Regional Model Performance Respective to Berkeley Earth Data | | | | (II) Regional Model Performance Respective to GISTEMP Data | | | |
|---|---|---|---|---|---|---|---|
| Model | Arctic | Antarctica | Tropics | Model | Arctic | Antarctica | Tropics |
| MMM | 0.0000496 | 0.0000889 | 0.0000167 | MMM | 0.000228 | 0.0000433 | 0.0000223 |
| AWI-CM-1-1-MR | 0.00018 | 0.00033 | 0.0000199 | AWI-CM-1-1-MR | 0.000338 | 0.000286 | 0.0000255 |
| BCC-CSM2-MR | 0.00069 | 0.000103 | 0.00000379 | BCC-CSM2-MR | 0.000908 | 0.000157 | 0.0000019 |
| EC-Earth3 | 0.00535 | 0.000362 | 0.0000707 | EC-Earth3 | 0.00616 | 0.000169 | 0.0000797 |
| FGOALS-g3 | 0.000165 | 0.0000474 | 0.00000458 | FGOALS-g3 | 0.000145 | 0.0000167 | 0.00000765 |
| GISS-E2-1-G | 0.0000408 | 0.000132 | 0.000033 | GISS-E2-1-G | 0.000209 | 0.0000527 | 0.0000425 |
| GFDL-CM4 | 0.000275 | 0.0000758 | 0.00000633 | GFDL-CM4 | 0.000631 | 0.0000678 | 0.00000805 |
| E3SM-1-1 | 0.00059 | 0.000102 | 0.0000225 | E3SM-1-1 | 0.0008 | 0.000155 | 0.0000256 |
| CESM2 | 0.000241 | 0.000281 | 0.0000172 | CESM2 | 0.000301 | 0.000209 | 0.0000223 |
| IPSL-CM6A-LR | 0.000076 | 0.000262 | 0.0000519 | IPSL-CM6A-LR | 0.0000385 | 0.000197 | 0.000063 |
| NorESM2-LM | 0.000507 | 0.0000596 | 0.0000474 | NorESM2-LM | 0.000623 | 0.0000223 | 0.0000526 |
| KIOST-ESM | 0.00019 | 0.000149 | 0.0000418 | KIOST-ESM | 0.000118 | 0.0000696 | 0.0000519 |

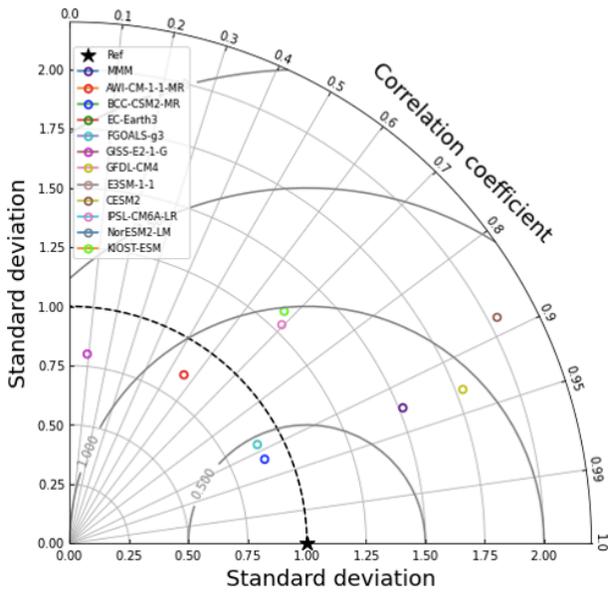
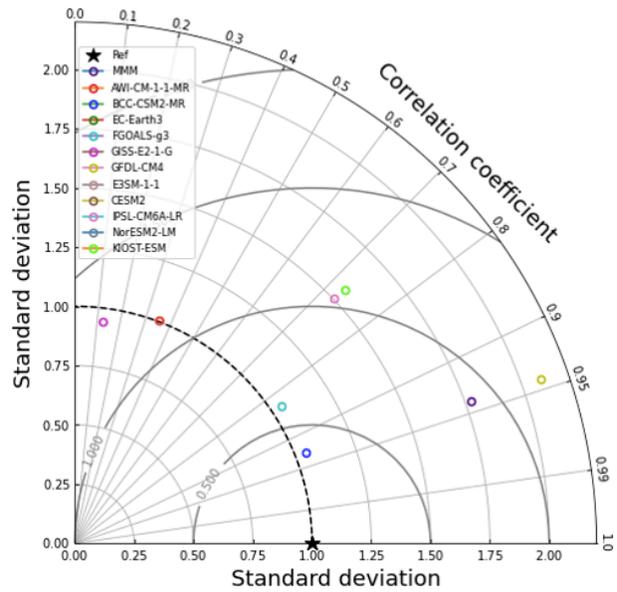

**Figure *S1***: Taylor diagrams evaluating CMIP6 models relative to the observations in the tropics.

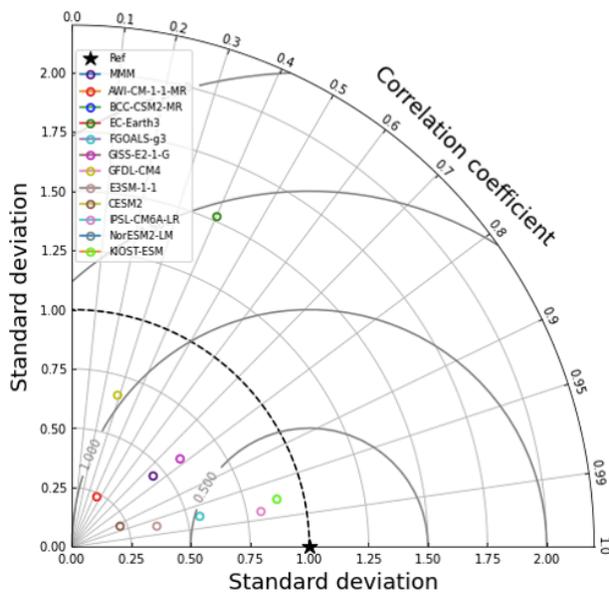
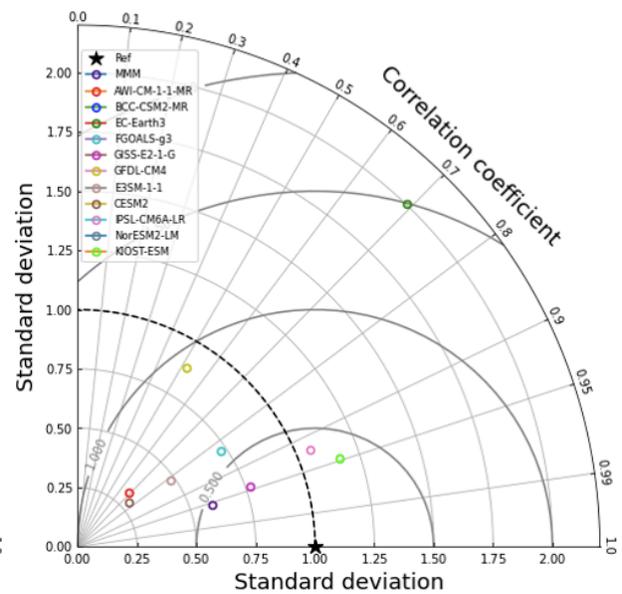

**Figure *S2***: Taylor diagrams evaluating CMIP6 models relative to the observations in Arctic.

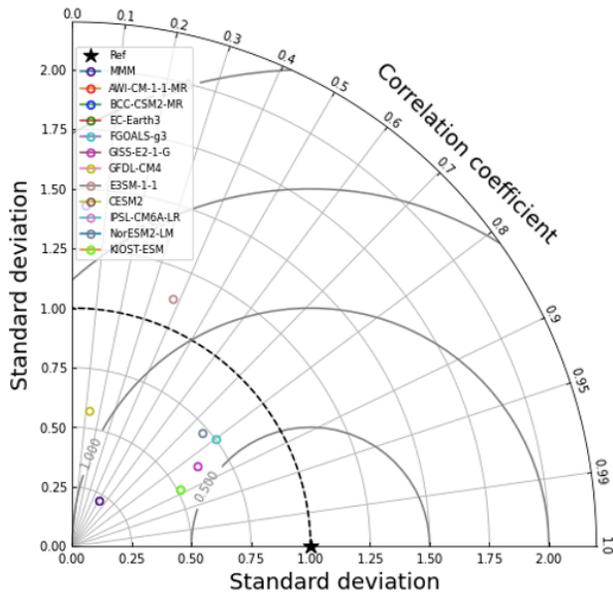 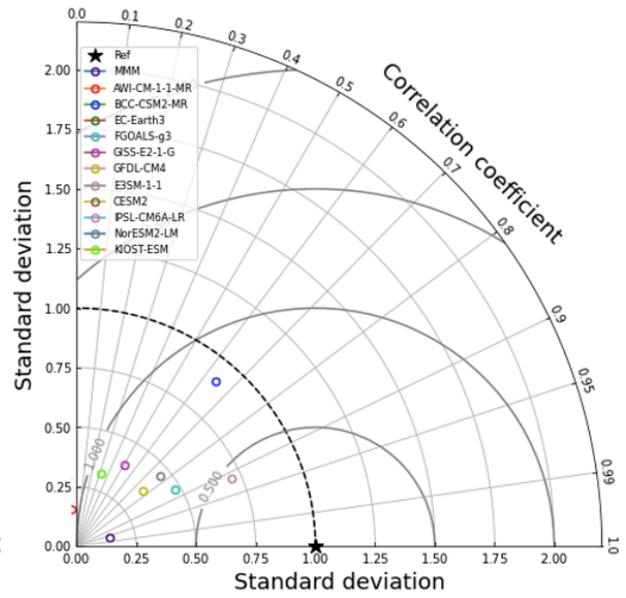

**Figure *S3***: Taylor diagrams evaluating CMIP6 models relative to the observations in Antarctica.